\pdfoutput=1
\documentclass[twocolumn,showpacs,amsmath,amssymb]{revtex4-1}
\usepackage{graphicx,color,hyperref}

\begin{document}

\title{Optical homodyne detection in view of joint probability distribution}
\author{Toru Kawakubo}
\email{kawakubo@nucleng.kyoto-u.ac.jp}
\author{Katsuji Yamamoto}
\affiliation{Department of Nuclear Engineering, Kyoto University,
Kyoto 606-8501, Japan}
\date{\today}

\begin{abstract}
Optical homodyne detection is examined
in view of joint probability distribution.
It is usually discussed that the relative phase between
independent laser fields are localized by photon-number measurements
in interference experiments such as homodyne detection.
This provides reasoning to use operationally coherent states
for laser fields in the description of homodyne detection
and optical quantum-state tomography.
Here, we elucidate these situations by considering
the joint probability distribution
and the invariance of homodyne detection
under the phase transformation of optical fields.
\end{abstract}


\maketitle

\section{Introduction}
\label{sec:introduction}

Laser technology plays essential roles
in a wide variety of fields in physics.
Since the first observation of interference fringes
between two independent laser fields \cite{Magyar1963},
it has been common to employ a coherent state \cite{Glauber1963}
\begin{equation}
| \alpha \rangle
= e^{-|\alpha|^2/2} \sum_{n=0}^\infty \frac{\alpha^n}{n!} | n \rangle
\label{eq:coherent-state}
\end{equation}
to describe the quantum state of a laser field,
which involves a coherent superposition of photon number states
$ | n \rangle $ with a definite phase in the complex amplitude $ \alpha $.
This quantum coherence provides indispensable resources
for quantum information and communication.
On the other hand, it is widely accepted \cite{Sargent1974,Walls1994}
that by considering the driving mechanism
the steady state of field inside a laser cavity
should be a mixed state as
\begin{equation}
\hat{\rho}_{|\alpha|} = e^{-|\alpha|^2} \sum_{n=0}^\infty
\frac{|\alpha|^{2n}}{n!} | n \rangle \langle n |
= \int \frac{d \varphi}{2 \pi}
| \alpha e^{i \varphi} \rangle \langle \alpha e^{i \varphi} | ,
\label{eq:mixed-state}
\end{equation}
which has no coherence, lacking a definite phase.
It is, however, shown by a numerical simulation \cite{Molmer1997}
that two cavity fields without definite phases even exhibit interference
when continuously monitored by photon detectors.
This provides a typical example for the apparent relevance
of the coherent state as the laser field.
Then, there have been a lot of debates
concerning the quantum state of laser and optical coherence
(see \cite{Molmer1997,Rudolph2001,vanEnk2001,
Sanders2003, Wiseman2004, Smolin2004, Cable2005, Neri2005,Bartlett2007},
and references therein).
The essential problem is whether the use of the coherent state
of Eq.\ (\ref{eq:coherent-state}) in various applications is valid or not
instead of the mixed state of Eq.\ (\ref{eq:mixed-state})
inside the laser cavity.

In quantum optics with the rotating wave approximation,
which is usually employed when describing matter-field interaction,
one can implement only the photon-number measurement,
without observing the absolute phases of fields.
Then, the U(1) invariance appears naturally in quantum optics
\cite{Sanders2003},
namely the photon number operator $ \hat{n} = \hat{a}^\dagger \hat{a} $
of each optical mode is invariant under the phase transformation,
$ \hat{a}^\dagger \to \hat{a}^\dagger e^{i \varphi} $
and $ \hat{a} \to \hat{a} e^{- i \varphi} $
for the creation and annihilation operators
$ \hat{a}^\dagger $ and $ \hat{a} $, respectively,
inducing $ | \alpha \rangle \to | \alpha e^{i \varphi} \rangle $
for the coherent state.
This phase transformation has intimate relation to the fact
that the phase of a single mode solely has no physical relevance,
i.e., there is no absolute reference frame for the optical phases
\cite{Bartlett2007}.
In this sense, it is trivial that there appears no significant difference
between the coherent state of Eq.\ (\ref{eq:coherent-state})
and the mixed state of Eq.\ (\ref{eq:mixed-state})
as long as U(1)-invariant operations and measurements
are performed starting only with a single-mode optical field.

The real issue to be clarified is rather
the interference between two independent optical fields,
which are mutually incoherent
without definite phases as seen in Eq.\ (\ref{eq:mixed-state}).
It has been discussed that photon-number measurements
induce localization of the relative phase
when two mutually incoherent fields interfere
\cite{Sanders2003,Cable2005,Neri2005}.
Actually, after many photons are detected in an interference experiment,
the relative phase is eventually localized around a certain value,
and the remaining state gets projected
to have a some definite relative phase.
Thus, the later measurement outcomes exhibit an interference pattern.

The aim of this paper is to elucidate these situations
in optical interference experiments,
where laser fields are treated operationally as coherent states.
Specifically, we consider homodyne detection and quantum-state tomography.
In order to illustrate the apparent relevance
for the use of coherent states,
we consider the joint probability distribution of the measurement outcomes
and the resultant empirical measure determining the quadrature distribution.
In this examination we adopt the proper description
of the output field of laser \cite{vanEnk2001}.
We also note that the homodyne detection
or generally photon-number detections are invariant
under the rotation of the phase frame over optical fields.

This paper is organized as follows.
In Sec.\ \ref{sec:homodyne}, we consider repeated homodyne detections
for independent signal and local oscillator fields
to see the localization of the relative phase.
In Sec.\ \ref{sec:joint-prob}, we introduce
the joint probability distribution of the outcomes
of the homodyne detections, and discuss the apparent relevance
to use the coherent state as the laser field
in the description of homodyne detection.
In Sec.\ \ref{sec:tomography}, we consider
the optical quantum-state tomography
based on these arguments on the homodyne detection,
and examine the quantum states of laser fields.
Sec.\ \ref{sec:summary} is devoted to summary.

\section{Phase localization by homodyne detections}
\label{sec:homodyne}

It has been argued \cite{Sanders2003,Cable2005,Neri2005}
that the relative phase gets localized by measurements
when two mutually incoherent fields interfere.
Here, we consider the homodyne detection
as a typical example of interference experiment
to discuss the phase localization, which provides a reason
why a coherent state may be adopted operationally
for the local oscillator (LO) field as the reference.

\begin{figure}
\includegraphics[width=.6\linewidth]{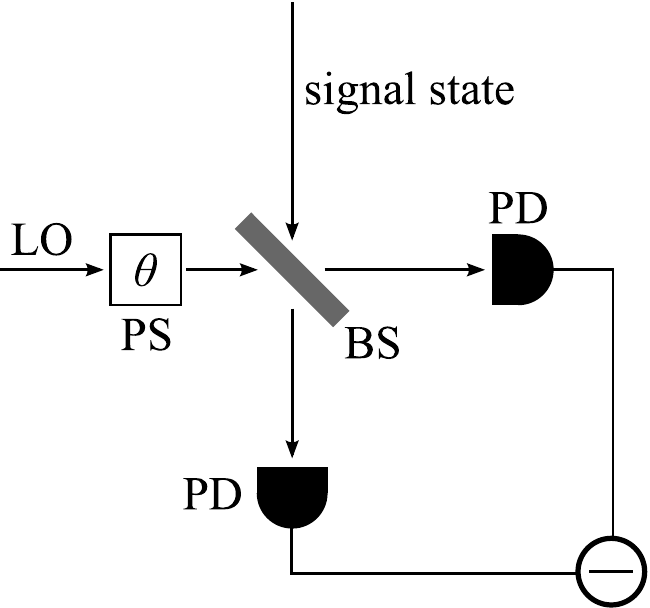}
\caption{Schematic diagram of homodyne detection.}
\label{fig:homodyne}
\end{figure}

The homodyne detection is a scheme to measure the field quadratures
and their probability distribution (see Fig.\ \ref{fig:homodyne}).
A signal field and a local oscillator are injected together
into a 50:50 beam splitter, and then the difference of the photon counts
in the output modes is measured.
The quadrature of the signal field is given
with a coherent state LO $ | \alpha \rangle$
($ \alpha \equiv e^{i \theta} | \alpha | $) by
\begin{equation}
\hat{x}_{\theta}
= ( \hat{a} e^{- i \theta} + \hat{a}^\dagger e^{i \theta} ) / \sqrt{2} ,
\label{eq:quad}
\end{equation}
which is approximately proportional to the photon-number difference
$ \Delta \hat{n} \simeq \sqrt{2} | \alpha | \hat{x}_{\theta} $
for $ |\alpha| \gg 1 $ \cite{Tyc2004}.
This quantity is sensitive to the relative phase between the signal and LO.

In usual homodyne experiments, the signal and LO are derived
from a common laser field to compensate the phase fluctuation.
In such a case, the laser field can be regarded as a coherent state
since its absolute phase, which is inherited equally by the signal and LO,
is irrelevant (or unobservable) in the homodyne detection
sensitive to the relative phase.
Instead, we here consider the case where the signal and LO
come from \emph{independent} sources, which are thus mutually incoherent.

We adopt an ideal continuous-wave (CW) laser as the LO.
The quantum state of the CW laser field is described properly
according to a formalism in Ref. \cite{vanEnk2001}.
[A pulsed-wave (PW) laser (not phase-locked one)
will also be considered in Sec.\ \ref{sec:tomography}
to discuss the optical quantum-state tomography
though the phase localization is unavailable for the PW case.]
Specifically, the output field of laser is separated
into a sequence of wave packet modes, each with the same duration.
By assuming the mixed state as given in Eq.\ (\ref{eq:mixed-state})
for the field inside the laser cavity
and the linear coupling between the modes inside and outside the cavity,
the sequence of $N$ output packets is described as
\begin{equation}
\hat{\rho}_\text{CW} = \int \frac{d \varphi}{2 \pi} P ( \varphi )
\left( | \alpha e^{i \varphi} \rangle
\langle \alpha e^{i \varphi} | \right)^{\otimes N} ,
\label{eq:CW}
\end{equation}
where $ \alpha > 0 $ is determined in terms of
the field intensity inside the cavity, the cavity leakage rate
and the packet duration.
This is a mixture of tensor products of coherent states
over their unknown phases.
Here, the phase distribution $ P( \varphi ) $
is introduced generally, which may be non-uniform,
while the Poissonian photon-number distribution is maintained.
As discussed in Ref. \cite{vanEnk2001},
this form for the output state of laser is exchangeable among the packets
to meet the quantum de Finetti theorem \cite{Hudson1976,Caves2002}.
The phase distribution changes apparently
as $ P ( \varphi ) \to P ( \varphi - \theta ) $
under the phase rotation
$ \mathcal{U}_\theta \hat{\rho}_\text{CW} \mathcal{U}_\theta^\dagger $.

We may take a tensor product of $ N $ packets
as $ \hat{\rho}^{\otimes N} $ for the signal field.
That is, many identical copies of $ \hat{\rho} $
are prepared and measured repeatedly in the homodyne detection.
Then, the initial state of the system is given by
\begin{equation}
\hat{\rho}^{\otimes N} \otimes \hat{\rho}_\text{CW}
= \int \frac{d \varphi}{2 \pi} P ( \varphi )
\left( \hat{\rho} \otimes | \alpha e^{i \varphi} \rangle
\langle \alpha e^{i \varphi} | \right)^{\otimes N} .
\label{eq:state-0}
\end{equation}
[Similar argument is also applicable
for a phase-mixture signal, as given in Eq.\ (\ref{eq:CW}),
as long as the relative phase between the signal and LO is concerned.
This case will be considered explicitly in Sec.\ \ref{sec:tomography}.]
After the first packet is measured, the remaining packets are projected to
\begin{equation}
\int \frac{d \varphi}{2 \pi} P ( \varphi ) q_\varphi(x)
\left( \hat{\rho} \otimes | \alpha e^{i \varphi} \rangle
\langle \alpha e^{i \varphi} | \right)^{\otimes (N-1)}
\label{eq:state-1}
\end{equation}
up to the normalization.
The quadrature value $x$ is determined
from the detected photon-number difference $\Delta n$ by
\begin{equation}
x = \Delta n / ( \sqrt{2} \alpha ) .
\label{eq:x-n}
\end{equation}
The quadrature distribution $ q_\varphi (x) $ for the signal $ \hat{\rho} $
with a pure coherent state LO $ | \alpha e^{i \varphi} \rangle $ is given by
\begin{equation}
\frac{q_\varphi (x) [ \hat{\rho} ]}{\sqrt{2} \alpha}
= \sum_{n-m=\Delta n} \langle n,m | \mathcal{B}
\left[ \hat{\rho} \otimes | \alpha e^{i \varphi} \rangle
\langle \alpha e^{i \varphi} | \right] | n,m \rangle ,
\label{eq:dist-def}
\end{equation}
where $ \mathcal{B} $ represents the unitary transformation
by the 50:50 beam splitter.  Specifically, for the strong LO field
we have approximately \cite{Tyc2004}
\begin{equation}
q_\varphi(x)
= \langle {\hat{x}_\varphi = x} | \hat{\rho} | {\hat{x}_\varphi = x} \rangle
\quad ( \alpha \to \infty ) ,
\label{eq:strong-approx}
\end{equation}
where $ | {\hat{x}_\varphi = x} \rangle $ is an eigenstate of the quadrature
$ \hat{x}_\varphi $ with an eigenvalue $ x $.

Repetition of $M$ detections with outcomes $ x_1, \ldots, x_M $
leads the state of the LO field to be
\begin{equation}
\int \frac{d \varphi}{2 \pi}
\left[ P ( \varphi ) \prod_{i=1}^M q_\varphi(x_i) \right]
\left( | \alpha e^{i \varphi} \rangle
\langle \alpha e^{i \varphi} | \right)^{\otimes (N-M)} .
\label{eq:state-M}
\end{equation}
Typically, for a coherent state signal
$ \hat{\rho} = | \beta \rangle \langle \beta | $ ($ \beta > 0 $)
we have the quadrature distribution ($ \alpha \to \infty $)
\begin{equation}
\label{eq:coherent-single}
q_\varphi(x) = \frac{1}{\sqrt{\pi}}
\exp \left[ -2 \beta^2 \left( \cos \varphi
- \frac{x}{\sqrt{2}\beta} \right)^2 \right] .
\end{equation}
Then, the phase distribution of the LO field after the $M$ detections
is modified from $ P ( \varphi ) $ by the factor
\begin{equation}
\prod_{i=1}^M q_\varphi(x_i)
\propto \left\{
\exp \left[ - 2 \beta^2 \left( \cos \varphi
- \frac{\bar{x}_M}{\sqrt{2}\beta} \right)^2 \right] \right \}^M ,
\label{eq:coherent-dist}
\end{equation}
where $ \bar{x}_M $ is the average over the $ M $ outcomes.
This provides a sharp Gaussian distribution
around $ \cos \varphi = \bar{x}_M / ( \sqrt{2} \beta ) $
with the standard deviation $ 1 / ( 2 \beta \sqrt{M} ) $ for $ M \gg 1 $.
That is, the phase of the LO field gets localized
to $ \varphi_0 = \pm \arccos ( \bar{x}_M / \sqrt{2} \beta ) $
by the repeated homodyne detections.
Generally, as long as $ q_\varphi(x) $ is non-uniform
with respect to the phase $ \varphi $ of the LO
(more precisely the relative phase between the signal and LO),
the phase localization takes place after a large number of detections
as (up to the normalization)
\begin{equation}
\prod_{i=1}^M q_\varphi (x_i)
\approx \delta ( \varphi - \varphi_0 ) .
\label{eq:localization}
\end{equation}
The phase localization is usually discussed
in the case of single measurement ($ M = 1 $)
with detection of large numbers of photons
under strong sources ($ \alpha , \beta \to \infty $)
\cite{Sanders2003,Cable2005,Neri2005}.
Here, we note that the phase localization takes place
by repeated measurements ($ M \gg 1 $)
even for the source packets with weak amplitudes
($ \alpha , \beta \sim 1 $).
This indeed provides a process of aligning the reference frames
between the signal and LO \cite{Bartlett2007}
by updating the relative phase according to the Bayesian rule
in Eq.\ (\ref{eq:state-M})
\cite{vanEnk2001,Buzek1998,Schack2001,Neri2005}.
The localized phase $ \varphi_0 $ may apparently take multiple values,
reflecting a specific symmetry of the signal state,
though they are physically equivalent.
For example, in the case of $ \hat{\rho} = | \beta \rangle \langle \beta | $
we have $ \varphi_0 = \pm \arccos ( \bar{x}_M / \sqrt{2} \beta ) $
for $ q_\varphi (x) = q_{- \varphi} (x) $
under the phase reflection $ \varphi \to - \varphi $.

Once the phase of the LO field is localized
to a particular value $ \varphi_0 $ in Eq.\ (\ref{eq:localization}),
the state of the LO field conditioned on the outcomes
$ x_1, \ldots , x_M $ gets projected as
\begin{equation}
\hat{\rho}_\text{CW}^{(x_1, \ldots , x_M)}
\approx \left( | \alpha e^{i \varphi_0} \rangle
\langle \alpha e^{i \varphi_0} | \right)^{\otimes (N-M)} .
\label{eq:final-state}
\end{equation}
Thus, we may conclude that the pure coherent state
$ | \alpha e^{i\varphi_0} \rangle $ is provided as the LO
for the subsequent detections
in the same way as the usual homodyne detection.
The apparent exception of the phase localization
is the case that the signal state $ \hat{\rho} $ is invariant
under the phase transformation,
including the number states and their mixtures.
Nevertheless, the quadrature distributions for such a state
with the pure coherent state LO
$ | \alpha e^{i \varphi} \rangle $ in Eq.\ (\ref{eq:coherent-state})
and the mixed state LO in Eq.\ (\ref{eq:mixed-state}) are identical
as $ q_\varphi(x) = q (x) $ independently of $ \varphi $.
Thus, even in this case without phase localization,
the laser field can be regarded as the coherent state.
This point will be clarified further
in view of the joint probability distribution in the following sections.

\section{Joint probability distribution in homodyne detections}
\label{sec:joint-prob}

We have seen that after repeated detections,
the state of the LO field turns into the product of coherent states
in Eq.\ (\ref{eq:final-state}) due to the phase localization.
This provides reasoning to use the coherent state
in the standard description of homodyne detection.
In order to get further understanding of this point
from a viewpoint of probability theory,
we here consider the joint probability distribution
of homodyne detections.

In quantum theory, measurements of a physical quantity
yield probabilistic outcomes.
Then, from the relative frequency of outcomes we can infer
the probability distribution for the physical quantity.
This argument is based on the assumption
that the  outcomes are \emph{independent and identically
distributed} ({i.i.d.}) in repeated measurements
for an ensemble of identically prepared quantum states.
Specifically, in the optical quantum-state tomography \cite{Lvovsky2009}
the quadrature distributions are determined from the outcomes
of homodyne detections.
In the standard description, a product of pure coherent states
$ \left( | \alpha e^{i \varphi_0} \rangle
\langle \alpha e^{i \varphi_0} | \right)^{\otimes N} $
with a common phase $ \varphi_0 $ is adopted as the LO packets
when the homodyne detections are performed repeatedly
for an ensemble of signal states as $ \hat{\rho}^{\otimes N} $.
Then, the joint probability distribution
of the outcomes $ x_1, \ldots, x_M $ is given by
\begin{equation}
p(x_1, \ldots, x_M) = \prod_{i=1}^M q_{\varphi_0} (x_i) ,
\label{eq:iid-prob}
\end{equation}
where $ q_{\varphi_0} (x) $ is the quadrature distribution
of the signal in Eq.\ (\ref{eq:dist-def}).
In this case the outcomes are really {i.i.d.},
namely they are obtained probabilistically according
to the product of identical quadrature distributions.
Thus, owing to the Glivenko-Cantelli theorem
the original distribution $ q_{\varphi_0} (x) $ is properly inferred
as the relative frequency of outcomes for $ M \to \infty $.

This argument for the standard homodyne detection
with the pure coherent state LO can be extended
for the case of the real output field of a CW laser
whose quantum state is the mixture as given in Eq.\ (\ref{eq:CW}).
The (unnormalized) state of the LO field after $M$ detections
is given in Eq.\ (\ref{eq:state-M}).
By tracing out the remaining packets
as $ \operatorname{Tr} \left[ ( | \alpha e^{i\varphi} \rangle
\langle \alpha e^{i\varphi} | )^{\otimes (N-M)} \right] = 1 $,
the joint probability distribution of the outcomes
is calculated as
\begin{equation}
p(x_1, \ldots , x_M) = \int \frac{d \varphi}{2 \pi} P ( \varphi )
\prod_{i=1}^M q_\varphi (x_i) .
\label{eq:joint-prob}
\end{equation}
Even in this extended case,
where the joint probability distribution in Eq.\ (\ref{eq:joint-prob})
appears as a phase-mixture of the {i.i.d.}\ products
in Eq.\ (\ref{eq:iid-prob}),
we can infer the original quadrature distributions
from the measurement outcomes as described in the following.

Consider a sequence of random real variables (measurement outcomes),
\begin{equation}
\tilde{x}_M \equiv (x_1,x_2, \ldots ,x_M) .
\end{equation}
The empirical measure $ \Lambda_{\tilde{x}_M} $,
or relative frequency of the $ M $ outcomes, is defined
as a probability measure on $\mathbb{R}$ by
\begin{equation}
\Lambda_{\tilde{x}_M}
= \frac{1}{M} \sum_{i=1}^M \delta_{x_i} ,
\label{eq:emp-meas}
\end{equation}
where $\delta_x$ denotes the Dirac measure on $\mathbb{R}$:
\begin{equation}
\delta_x(A) = \begin{cases}
1 & \text{if $x\in A\subset\mathbb{R}$,} \\
0 & \text{otherwise.} \end{cases}
\end{equation}
That is, if the number of $x_i$'s which have values in $A$ is $k$,
then $ \Lambda_{\tilde{x}_M} (A) = k/M $.
In the {i.i.d.}\ case of Eq.\ (\ref{eq:iid-prob}),
the Glivenko-Cantelli theorem ensures
that the empirical measure $ \Lambda_{\tilde{x}_M} $ converges
to the original distribution $ q_{\varphi_0}(x) $ for $ M \to \infty $.
As for the actual homodyne detection with the LO of a CW laser field,
the joint probability distribution in Eq.\ (\ref{eq:joint-prob})
represents a mixture of the {i.i.d.}\ variables (or {i.i.d.}'s shortly).
Even in this case, by repeating the detection many times ($  M \gg 1 $)
the empirical measure provides the quadrature distribution
with a certain random phase $ \varphi_0 $,
\begin{equation}
\lim_{M \to \infty}
\Lambda_{\tilde{x}_M} (x) = q_{\varphi_0} (x)
\label{eq:Lambda-q}
\end{equation}
(see Ref. \cite{Aldous1985} for the mathematical details).
This implies that the outcomes $ x_1, \ldots, x_M $
appear as if they were {i.i.d.},
in the same way as the case with the pure coherent state LO.
Therefore, in each sequence of homodyne detections
we may regard the LO of the CW laser field
as \emph{a coherent state} $ | \alpha e^{i \varphi_0} \rangle $
while the phase $ \varphi_0 $ is determined \emph{a posteriori}
by the localization.

We have made numerical simulations for the joint probability distributions
of homodyne detections, confirming Eq.\ (\ref{eq:Lambda-q}).
A sequence of $ M $ outcomes are obtained
according to Eq.\ (\ref{eq:joint-prob})
representing the mixture of {i.i.d.}'s.
Specifically, the $i$-th outcome $ x_i $ is generated
under the conditional probability distribution,
\begin{eqnarray}
p(x_i | x_1, \ldots ,x_{i-1})
&=& \frac{p(x_1, \ldots ,x_i)}{p(x_1, \ldots ,x_{i-1})}
\nonumber \\
&=& \int \frac{d \varphi}{2 \pi} P_{(x_1, \ldots ,x_{i-1})} ( \varphi )
q_\varphi (x_i)
\end{eqnarray}
with
\begin{equation}
p(x_1, \ldots ,x_i)
= \int d x_{i+1} \cdots d x_M p(x_1, \ldots ,x_M) .
\end{equation}
(A similar analysis is made for the spatial interference
of Bose-Einstein condensates \cite{Javanainen1996}.)
Here, the phase distribution is updated by the Bayesian rule
\cite{vanEnk2001,Buzek1998,Schack2001,Neri2005}
upon the preceding quadrature outcomes $ x_1, \ldots ,x_{i-1} $ as
\begin{equation}
P_{(x_1, \ldots , x_{i-1})} ( \varphi )
= \prod_{j=1}^{i-1} q_\varphi (x_j)
\left/ \int \frac{d \varphi}{2 \pi} \prod_{j=1}^{i-1} q_\varphi (x_j)
\right.
\label{eq:P-update}
\end{equation}
with the U(1)-invariant initial $ P ( \varphi ) = 1 $ for definiteness.
The empirical measure $ \Lambda_{\tilde{x}_M} (x) $
is then calculated from the $ M $ outcomes with Eq.\ (\ref{eq:emp-meas}).
In this numerical analysis, the original quadrature distribution
$ q_\varphi(x) $ is calculated precisely from Eq.\ (\ref{eq:dist-def})
without taking the limit of strong laser intensity.
Statistically, a large number of detections should be made
to infer the quadrature distribution.
Thus, we realize in Eq.\ (\ref{eq:P-update})
that in the early portion of the $ M $ detections ($ M \gg 1 $)
the phase $ \varphi $ is almost localized to $ \varphi_0 $
providing Eq.\ (\ref{eq:Lambda-q}).

\begin{figure}
\includegraphics{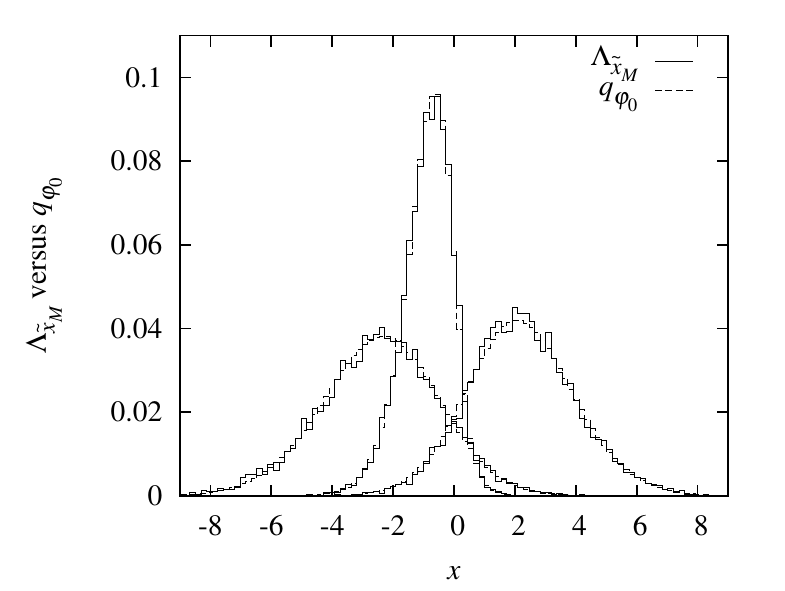}
\caption{Typical results of the empirical measure
$ \Lambda_{\tilde{x}_M} (x) $
(solid lines) are shown for the squeezed state signal
$ \hat{\rho} = | r , \beta \rangle \langle r , \beta | $,
where $ M=10000 $ outcomes are generated
for each operation of repeated homodyne detections.
The parameters for the LO and signal is taken as
$ \alpha = \sqrt{15} $, $ \beta e^{-r} = {\sqrt 3} $ and $ r = - 1 $.
The resolution of the quadrature
is given by $ \Delta x = 1 / ( \sqrt{2} \alpha ) = 1 / \sqrt{30} $.
The resultant random phases $ \varphi_0 $
are estimated from the average $ \bar{x}_M $ of the outcomes
as $ \varphi_0 = 3.05\,\text{rad} $, $1.94\,\text{rad} $, $0.43\,\text{rad} $
from the left to right.
The quadrature distributions $ q_{\varphi_0} (x) $
(dashed lines) are also plotted for these values of $ \varphi_0 $,
showing good agreement with $ \Lambda_{\tilde{x}_M} (x)$.}
\label{fig:emp-meas}
\end{figure}

Fig.\ \ref{fig:emp-meas} shows typical results
of the empirical measure $ \Lambda_{\tilde{x}_M} (x) $
(solid lines) for the squeezed state signal
$ \hat{\rho} = | r , \beta \rangle \langle r , \beta | $,
where $ M=10000 $ outcomes are generated
for each operation of repeated homodyne detections.
The parameters for the LO and signal is taken as
$ \alpha = \sqrt{15} $, $ \beta e^{-r} = {\sqrt 3} $ and $ r = - 1 $.
The resolution of the quadrature $ x $
is given by $ \Delta x = 1 / ( \sqrt{2} \alpha ) $
in Eq.\ (\ref{eq:x-n}) with $ \Delta n = 1 $.
We note that the expectation value of the quadrature
is calculated with $ q_{\varphi} (x) $
as $ \sqrt{2} \beta e^{-r} \cos \varphi $
independently of $ \alpha > 0 $ for the coherent state LO.
Since it should agree with the average $ \bar{x}_M $ of the outcomes
for $ M \to \infty $, the resultant random phases $ \varphi_0 $
are estimated as
$ \varphi_0 = 3.05\,\text{rad} $, $1.94\,\text{rad} $, $0.43\,\text{rad} $
from the left to right in Fig.\ \ref{fig:emp-meas}.
Then, the quadrature distributions $ q_{\varphi_0} (x) $
(dashed lines) are plotted for these values of $ \varphi_0 $ for comparison.
These results really show good agreement
of $ \Lambda_{\tilde{x}_M} (x) $ and $ q_{\varphi_0} (x) $,
as expected in Eq.\ (\ref{eq:Lambda-q}).

\section{Optical quantum-state tomography}
\label{sec:tomography}

In the view of joint probability distribution
for homodyne detection, as described in the previous section,
we now consider the optical quantum-state tomography,
and discuss the quantum state of a laser field.
The signal state (e.g., Wigner function) is reconstructed
from the quadrature distributions $ q_\theta (x) $
for various phase shifts $ \theta $,
which are obtained as the empirical measures
from the outcomes of homodyne detections.
The change of $ \theta $ is realized
by applying a phase shifter on the LO.

\subsection{Tomography with a common source oscillator}

We first consider the usual setup for optical tomography,
where the signal and LO are supplied by splitting a single oscillator,
as done in many actual experiments.
The output state of a CW laser for the original oscillator
($ \alpha , \beta > 0 $) is given as
\begin{equation}
\hat{\rho}_\text{CW}^0
= \int \frac{d \varphi}{2 \pi} P ( \varphi )
\left[ | ( \alpha + \beta ) e^{i \varphi} \rangle
\langle ( \alpha+ \beta )  e^{i \varphi} | \right]^{\otimes N} .
\end{equation}
The signal and LO, which share the common random phase $ \varphi $,
are derived from this laser field as
\begin{equation}
\hat{\rho}_\text{CW}^{\text{SL}}
= \int \frac{d \varphi}{2 \pi} P ( \varphi )
\left[ \hat{\rho} ( \varphi ) \otimes
| \alpha e^{i ( \varphi + \theta )} \rangle
\langle \alpha e^{i ( \varphi + \theta )} | \right]^{\otimes N}
\end{equation}
with
\begin{equation}
\hat{\rho} ( \varphi ) = \mathcal{E}
| \beta e^{i \varphi} \rangle \langle \beta e^{i \varphi} |
\mathcal{E}^\dagger
= \mathcal{U}_\varphi \hat{\rho} (0) \mathcal{U}_\varphi^\dagger .
\end{equation}
Here, the phase of the LO is shifted by $ \theta $,
and the operation $ \mathcal{E} $ such as squeezing
is applied for the signal, which commutes with the phase transformation
$ \mathcal{U}_\varphi $.
We see below that this setup reproduces
the standard description of homodyne tomography
with a pure coherent state $ | \alpha \rangle $ as the LO.
The joint probability distribution
of the quadrature outcomes
$ ( x_1, \ldots, x_M ) \equiv \tilde{x}_M $ ($ M \gg 1 $)
for $ M $ packets is calculated as
\begin{eqnarray}
p( \tilde{x}_M ) [ \hat{\rho}_\text{CW}^{\text{SL}} ]
&=& \int \frac{d \varphi}{2 \pi} P ( \varphi )
\prod_{i=1}^M q_{\varphi + \theta} (x_i) [ \hat{\rho} ( \varphi ) ]
\nonumber \\
&=& \prod_{i=1}^M q_{\theta} (x_i) [ \hat{\rho} (0) ] .
\label{eq:pCWSL}
\end{eqnarray}
Here, we have considered the fact
that the homodyne detection is an invariant operation
under the simultaneous phase rotation $ \mathcal{U}_\varphi $
for the signal and LO as
$ | \alpha e^{i ( \varphi + \theta )} \rangle
\to | \alpha e^{i \theta} \rangle $
and $ | \beta e^{i \varphi} \rangle \to | \beta \rangle $,
which implies
\begin{equation}
q_{\varphi + \theta} (x) [ \hat{\rho} ( \varphi ) ]
= q_{\theta} (x) [ \hat{\rho} (0) ] ,
\end{equation}
i.e., it is sensitive only to the relative phase $ \theta $.
This $ p( \tilde{x}_M ) $ turns out to be independent
of the phase distribution $ P( \varphi ) $ for the LO,
that is the possible U(1) violation in the laser field is not observable
in this scheme.
The quadrature outcomes are {i.i.d.}\ in Eq.\ (\ref{eq:pCWSL}),
and the LO appears as if it is the coherent state
$ | \alpha e^{i \theta} \rangle $ with $ \varphi = 0 $.
Thus, by repeating independently the sequence of $ M $ detections
on $ \hat{\rho}_\text{CW}^{\text{SL}} $ from the common source
with the varying phase shift $ \theta $ for the LO,
the set of quadrature distributions $ q_{\theta} (x) [ \hat{\rho} (0) ] $
is obtained to reconstruct the signal state through tomography as
\begin{equation}
\hat{\rho}_\text{rec} = \hat{\rho} (0)
= \mathcal{E} ( | \beta \rangle \langle \beta | ) \mathcal{E}^\dagger ,
\end{equation}
which is irrespective of the unknown phase $ \varphi $.

Alternatively, we may  adopt a PW laser for the original oscillator,
providing $ N $ copies of a phase-mixture of coherent states,
\begin{equation}
\hat{\rho}_\text{PW}^0
= \left[ \int \frac{d \varphi}{2 \pi} P ( \varphi )
 | ( \alpha + \beta ) e^{i \varphi} \rangle
\langle ( \alpha+ \beta ) e^{i \varphi} | \right]^{\otimes N} .
\end{equation}
The combination of signal and LO is derived as
\begin{equation}
\hat{\rho}_\text{PW}^\text{SL}
= \left[ \int \frac{d \varphi}{2 \pi} P ( \varphi )
\hat{\rho} ( \varphi ) \otimes
| \alpha e^{i ( \varphi + \theta )} \rangle
\langle \alpha e^{i ( \varphi + \theta )} | \right]^{\otimes N} .
\label{eq:pPWSL}
\end{equation}
Then, the same joint probability distribution is obtained
as Eq.\ (\ref{eq:pCWSL}) for the CW case,
\begin{eqnarray}
p( \tilde{x}_M ) [ \hat{\rho}_\text{PW}^\text{SL} ]
&=& \prod_{i=1}^M \int \frac{d \varphi_i}{2 \pi} P ( \varphi_i )
q_{\varphi_i + \theta} (x_i) [ \hat{\rho} ( \varphi_i ) ]
\nonumber \\
&=& \prod_{i=1}^M q_{\theta} (x_i)  [ \hat{\rho} (0) ] ,
\end{eqnarray}
providing again the reconstruction of the signal state as
$ \hat{\rho} (0)
= \mathcal{E} ( | \beta \rangle \langle \beta | ) \mathcal{E}^\dagger $.
Therefore, as long as the common laser field is used for the signal and LO,
we find no actual difference between the CW and PW cases.
In either case, the use of a pure coherent state as the LO is relevant
for the standard description of the optical quantum-state tomography,
without need to discuss the phase localization.
The tomography with the common source just characterizes
the process given by the operation $ \mathcal{E} $
rather than the signal state \cite{Sanders2003}.

\subsection{Tomography with independent signal and LO}

We next consider the case that the signal and LO
are prepared independently,
which may be more faithful in the sense of tomography
to reconstruct an ``unknown'' quantum state.
The LO is supplied with the output state $ \hat{\rho}_\text{CW} $
of a CW laser as given in Eq.\ (\ref{eq:CW}).
An ensemble of repeatedly prepared identical states for the signal
may be given generally as
\begin{equation}
\hat{\rho}_\text{S}
= \int \frac{d \varphi^\prime}{2 \pi} P_\text{S} ( \varphi^\prime )
\hat{\rho} ( \varphi^\prime )^{\otimes N} ,
\label{eq:rhoS}
\end{equation}
where
\begin{equation}
\hat{\rho} ( \varphi^\prime )
\equiv \mathcal{U}_{\varphi^\prime} \hat{\rho} (0)
\mathcal{U}_{\varphi^\prime}^\dagger
\end{equation}
with certain $ \hat{\rho} (0) $.
The phase distributions
$ P ( \varphi ) $ and $ P_\text{S} ( \varphi^\prime ) $
(with period $ 2 \pi $) may not be invariant
under the rotation of phase frame.
In the case of a U(1)-invariant state
$ \hat{\rho}(0) = \hat{\rho} ( \varphi^\prime ) $
for any $ \varphi^\prime $, namely a mixture of number states
\cite{Sanders2003},
we simply have $ \hat{\rho}_\text{S} = \hat{\rho} (0)^{\otimes N} $
without $ P_\text{S} ( \varphi^\prime ) $.

The joint probability distribution of homodyne detections is calculated as
\begin{equation}
p( \tilde{x}_M ) [ \hat{\rho}_\text{S} \otimes \hat{\rho}_\text{CW} ]
= \int \frac{d \varphi}{2 \pi} \bar{P}_\text{S} ( \varphi )
\prod_{i=1}^M q_{\varphi}^{(0)} (x_i)
\label{eq:pxMCW}
\end{equation}
with a convoluted phase distribution
\begin{equation}
\bar{P}_\text{S} ( \varphi ) = \int \frac{d \varphi^\prime}{2 \pi}
P_\text{S} ( \varphi^\prime ) P ( \varphi + \varphi^\prime ) .
\label{eq:PSbar}
\end{equation}
Here, we have considered the invariance of homodyne detection
under the phase transformation, implying the relation
for the quadrature distributions as
\begin{equation}
q_\varphi (x) [ \hat{\rho} ( \varphi^\prime ) ]
\equiv q_\varphi^{( \varphi^\prime )} (x)
= q_{\varphi - \varphi^\prime}^{(0)} (x)
\label{eq:U(1)relation}
\end{equation}
with the periodicity $ q_{\varphi + 2 \pi} (x) = q_\varphi (x) $.
We find that this $ p( \tilde{x}_M ) $ represents a mixture of {i.i.d.}'s,
as discussed in Sec.\ \ref{sec:joint-prob}.
Then, a quadrature distribution $ q_{\varphi_0}^{(0)} (x) $ is obtained
as the empirical measure with the probability distribution
$ \bar{P}_\text{S} ( \varphi_0 ) $ for the random phase $ \varphi_0 $.
Here, we note that the U(1)-invariant LO with $ P ( \varphi ) = 1 $
provides $ \bar{P}_\text{S} ( \varphi ) = 1 $,
irrespective of any $ P_\text{S} ( \varphi ) $ for the original signal.
Contrarily, if any deviation of $ \bar{P}_\text{S} ( \varphi ) $
from the uniform distribution is found
for the various values of $ \varphi = \varphi_0 $
in experiments (provided $ \varphi_0 $ is determined in a certain situation,
e.g., from the average of outcomes
for the coherent or squeezed state signal), that is
\begin{equation}
\bar{P}_\text{S} ( \varphi ) \neq 1
\rightarrow P ( \varphi ) , P_\text{S} ( \varphi ) \neq 1 ,
\end{equation}
then it might indicate the violation of U(1) symmetry,
or the presence of some implicit phase reference
common to the signal and LO.

The measurement of the $ M $-packet sequence may be repeated independently.
Then, the quadrature distributions $ q_{\varphi_0}^{(0)} (x) $
are obtained with various random phases $ \varphi_0 $.
It is, however, impossible in general
to know the actual values of $ \varphi_0 $
without some prior knowledge about the signal state.
These unknown random phases $ \varphi_0 $ for $ q_{\varphi_0}^{(0)} (x) $
thus can not substitute for the phase shift $ \theta $ of the LO
in tomography.
The phase shift $ \theta $ of the LO
in each of the independent $ M $-packet sequences is actually ineffective
since it is hidden in the random phase $ \varphi_0 $.

Instead, in order to realize effectively the phase shift of the LO,
we should extend the single $ M $-packet sequence
to $ K \times M $-packet sequences in a single operation of tomography:
\begin{equation}
\hat{\rho}_\text{CW}
\rightarrow \int \frac{d \varphi}{2 \pi} P ( \varphi )
\prod_{k=1}^K \left[
\left( | \alpha e^{i ( \varphi + \theta_k )} \rangle
\langle \alpha e^{i ( \varphi + \theta_k )} | \right)^{\otimes M} \right] ,
\end{equation}
where the phase shift $ \theta_k $ is applied for the LO
in each $ M $-packet sequence.
Then, the joint probability distribution is given as
\begin{equation}
p( \tilde{x}_M^{(1)} , \ldots , \tilde{x}_M^{(K)} )
= \int \frac{d \varphi}{2 \pi} \bar{P}_\text{S} ( \varphi )
\prod_{k=1}^K \left[
\prod_{i=1}^M q_{\varphi + \theta_k}^{(0)} (x_i) \right] .
\end{equation}
This provides the sequence of empirical measures upon homodyne detections,
determining the quadrature distributions for tomography
with varying phases $ \theta_k $ ($ M \gg 1 $):
\begin{equation}
\Lambda_{\tilde{x}_M^{(1)}} = q_{\varphi_0 + \theta_1}^{(0)} (x) ,
\ldots , \Lambda_{\tilde{x}_M^{(K)}} = q_{\varphi_0 + \theta_K}^{(0)} (x) ,
\end{equation}
where the original phase of the LO is fixed to a certain value $ \varphi_0 $
according to the localization.

Provided there is no way to know the value of $ \varphi_0 $,
we may set $ \varphi_0 = 0 $ operationally
(as a convenient choice of the phase frame),
or consider the relation
\begin{equation}
q_{\varphi_0 + \theta_k}^{(0)} (x) = q_{\theta_k}^{( - \varphi_0 )} (x) .
\end{equation}
Then, the set of quadrature distributions
$ q_{\theta_k}^{( - \varphi_0 )} (x) $
for the pure coherent states LO $ | \alpha e^{i \theta_k} \rangle $
with the phase shifts $ \theta_k $ ($ \alpha > 0 $ and $ \varphi = 0 $)
provides the tomographic reconstruction as
\begin{equation}
\hat{\rho}_\text{rec} = \hat{\rho} ( - \varphi_0 ) .
\end{equation}
In each operation of tomography, the reconstructed state
$ \hat{\rho} ( - \varphi_0 ) $ appears probabilistically
as a random rotation of $ \hat{\rho} (0) $.
Due to the lack of the absolute phase reference, however,
$ \hat{\rho} (0) $ and $ \hat{\rho} ( - \varphi_0 ) $
should be regarded equivalent,
and the ensemble of signal states is properly inferred
as $ \hat{\rho}_\text{S} $ in Eq.\ (\ref{eq:rhoS})
with $ \bar{P}_\text{S} ( \varphi^\prime ) $.
These arguments illustrate the actual relevance
for the use of the pure coherent state as the LO
in the description of optical quantum-state homodyne tomography.
The random phases $ \varphi_0 $ and their distribution
$ \bar{P}_\text{S} ( \varphi_0 ) $ may be estimated relatively
by comparing the rotations for the resultant Wigner functions
obtained from many runs of tomography for the same $ \hat{\rho}_\text{S} $.
Note, however, that $ \bar{P}_\text{S} ( \varphi_0 ) = 1 $
for the U(1)-invariant LO with $ P ( \varphi ) = 1 $,
irrespective of the actual $ P_\text{S} ( \varphi^\prime ) $.

\subsection{CW field versus PW field}

We also consider the case that the quantum state of LO
is given by a product of mixed states,
which may be prepared with a simple PW laser (not a phase-locked one):
\begin{equation}
\hat{\rho}_\text{PW}
= \left[ \int \frac{d \varphi}{2 \pi} P ( \varphi )
| \alpha e^{i \varphi} \rangle \langle \alpha e^{i \varphi} |
\right]^{\otimes N} .
\label{eq:PW}
\end{equation}
In the PW case as the LO,
the joint probability distribution is calculated
for the signal state in Eq.\ (\ref{eq:rhoS}) as
\begin{equation}
p( \tilde{x}_M )
[ \hat{\rho}_\text{S} \otimes \hat{\rho}_\text{PW} ]
= \int \frac{d \varphi^\prime}{2 \pi} P_\text{S} ( \varphi^\prime )
\prod_{i=1}^M \bar{q}_{\varphi^\prime} (x_i) ,
\end{equation}
where
\begin{equation}
\bar{q}_{\varphi^\prime} (x)
\equiv \int \frac{d \varphi}{2 \pi} P ( \varphi )
q_{\varphi^\prime}^{( \varphi^\prime - \varphi + \varphi^\prime )} (x) .
\label{eq:q-bar}
\end{equation}
with the relation $ q_\varphi^{( \varphi^\prime )} (x)
= q_{\varphi^\prime}^{( \varphi^\prime - \varphi + \varphi^\prime )} (x) $
under the phase rotation $ \mathcal{U}_{- \varphi + \varphi^\prime} $.
This smeared quadrature distribution $ \bar{q}_{\varphi^\prime} (x) $
is reproduced with the LO state $ | \alpha e^{i \varphi^\prime} \rangle $
for the signal of a phase-mixed state
\begin{equation}
\hat{\rho}_\text{mix}
= \int \frac{d \varphi}{2 \pi} P ( \varphi )
\hat{\rho} ( \varphi^\prime - \varphi + \varphi^\prime ) ,
\end{equation}
which generally does not coincide with $ \hat{\rho} (0) $
or its phase-rotation $ \hat{\rho} ( \varphi^\prime ) $,
except for the U(1)-invariant $ \hat{\rho} (0) $.
Thus, we find that the optical quantum-state tomography
does not work rightly by using the PW field $ \hat{\rho}_\text{PW} $
in Eq.\ (\ref{eq:PW}) as the independent LO.
The phase-mixture of product coherent states
$ \hat{\rho}_\text{CW} $  in Eq.\ (\ref{eq:CW}),
which is derived from a CW laser, is required
for the successful tomography.
As an interesting case, we may implement the homodyne tomography
for the CW and PW fields with the independent CW field as the LO.
Then, we will obtain in the reconstruction
the coherent state in Eq.\ (\ref{eq:coherent-state}) for the CW signal,
and the mixed state in Eq.\ (\ref{eq:mixed-state}) for the PW signal,
respectively. In this way,
we can distinguish the quantum states of laser fields.

\section{Summary}
\label{sec:summary}

We have examined the repeated optical homodyne detections
and quantum-state tomography in view of the joint probability distribution
of the measurement outcomes.
By adopting the real output state of a CW laser as the LO,
which is independent of the signal field,
the joint probability distribution represents a mixture of {i.i.d.}'s.
Then, the original quadrature distribution of the signal
is obtained as the empirical measure, or relative frequency of the outcomes,
with a random phase for the coherent state LO determined a posteriori
by the phase localization according to the Bayesian rule.
This justifies the operational use of the coherent state as the LO
in the standard description of homodyne detection and tomography.
We have also discussed that the quantum states of CW and PW lasers
are distinguishable by the quantum-state tomography
with the independent CW field as the LO.
That is, the CW and PW lasers will appear
as the coherent state and the mixed state, respectively.
On the other hand, both of them will be recognized as the coherent state
indistinguishably if the tomography is implemented
with the signal and LO derived from the common source oscillator,
as usually made in optical experiments.

\begin{acknowledgments}
We thank K. Fujii for valuable discussions.
T. K. was supported by the JSPS Grant No.\ 22.1355.
\end{acknowledgments}

\providecommand \doibase [0]{http://dx.doi.org/}%
\providecommand \Doi[1]{\href{\doibase#1}}%
\providecommand \Eprint[0]{\href }%

\end{document}